\documentclass[twocolumn,aps,prd,epsf]{revtex4}
\usepackage{epsfig}

\begin{document}

\title{Prediction of the masses and decay processes of strange, charmed 
and bottomed pentaquarks from the linear molecular crypto-heptaquark model.}
\author{P. Bicudo}
\email{bicudo@ist.utl.pt}
\affiliation{Dep. F\'{\i}sica and CFIF, Instituto Superior T\'ecnico,
Av. Rovisco Pais, 1049-001 Lisboa, Portugal}
\begin{abstract}
In this paper the masses and decay processes of several new strange, 
charmed and bottomed exotic pentaquarks are predicted. 
Multiquarks are studied microscopically in a standard quark model. 
In pure ground-state pentaquarks the short-range interaction is computed 
and it is shown to be repulsive. The long-range and medium-range 
interactions are not expected to provide sufficient attraction.
An additional quark-antiquark pair is then considered, and this is 
suggested to produce a narrow linear molecular system. The quarks 
assemble in three hadronic clusters, and the central hadron provides 
stability. The possible crypto-heptaquark hadrons with exotic pentaquark 
flavours, with any number of strange, charmed and bottomed quarks,  
are listed. Several new exotics may still be observed.
\end{abstract}
\maketitle

\section{\label{sec:intro}introduction}

\par
In this paper the masses and decay processes of several new strange, charmed 
or bottomed exotic pentaquarks are predicted in the linear-molecule heptaquark
approach. 

Exotic multiquarks are expected since the early
works of Jaffe
\cite{Jaffe},
and the masses and decays in the SU(3) exotic anti-decuplet were first 
predicted within the chiral soliton model
\cite{Diakonov1}. 
The pentaquarks have been revived by several searches of the
$\Theta^+$(1540) 
\cite{Nakano,Barmin,Stepanyan,Barth,Asratyan,Kubarovsky,Airapetian,Juengst,
Aleev,Bai,Abdel-Bary,Knopfle,Aslanyan,Chekanov,Pinkenburg,Troyan,Raducci,
Antipov:2004jz,Abt:2004tz,Longo:2004gd,Aubert:2005qi,Wang:2005fc,Aslanyan:2005gn,
Hicks:2005gp,Battaglieri:2005er,
Aleev:2004sa,Aleev:2005yq,Kubarovsky:2005sk,Nakano:2005,
Kabana:2004hh,Huang:2005nk}, 
first discovered at LEPS
\cite{Nakano},
and by recent searches
of the $\Xi^{--}$(1860) and of the $D^{*-}p$(3100), respectively at NA49
\cite{Alt,Fischer,Price,Airapetian:2004mi,Chekanov:2005at,
Christian:2005kh}
and at H1
\cite{H1,Chekanov:2004qm,Zavertyaev:2005yf,Link:2005ti}. 
Presently the number of negative experiments is larger than the number
of positive experiments. For example the recent CLAS
higher statistics analysis contradicts previous CLAS
and ELSA experiments. Nevertheless the positive experiments don't allow us to
exclude the pentaquarks, but they show that the pentaquarks can only be produced
in certain processes. In particular, recent new experiments, by SVD-2 and
LEPS
\cite{Aleev:2004sa,Aleev:2005yq,Kubarovsky:2005sk,Nakano:2005}
 continue to confirm the $\Theta^+$ pentaquark. A new $\Theta^{++}$ 
pentaquark was also discovered recently at STAR
\cite{Kabana:2004hh,Huang:2005nk}, 
which suggests that the
pentaquark may have Isospin 1 or 2
\cite{Page}. 
If this is confirmed, the
theoretical pentaquark models predicting isospin 0 for the $\Theta$ would be 
ruled out. Experimental upgrades are already programmed to further scan the 
pentaquarks. Pentaquark structures
have also been studied in the lattice
\cite{Nemoto:2003ft,Csikor:2003ng,Sasaki:2003gi,Chiu:2004gg,Doi:2004jp,
Okiharu:2004wy,Alexandrou:2004ak,Ishii:2004qe,Okiharu:2004ve,
Ishii:2005zd,Lasscock:2005tt,Takahashi:2005uk,Ishii:2005vc,Basak:2005aq}.
In  the range of the $\Theta^+$ experimental mass, quenched lattice simulations 
only identify a parity - system, compatible with an open Kaon-nucleon channel. 
Pentaquarks motivate this effort, because they may be the first exotic 
hadrons ever discovered, with quantum numbers that cannot be interpreted
as a quark-plus-antiquark meson or as a three-quark baryon. Importantly, the observed 
pentaquarks have an extremely narrow decay width, two orders of magnitude smaller
than normal hadronic decay widths.

It is also remarkable that many different models presently dispute the interpretation 
of these multiquarks. It is well known that the pentaquark cannot be a simple five 
quark state in the groundstate, because it would freely recombine and decay in a Kaon 
and a nucleon, resulting in resonance with a very broad decay width. Thus all plausible 
models essentially propose an excitation which limits the decay width of the pentaquark. 
Because the models of non-pertubative hadronic physics are not yet sufficiently accurate
or properly calibrated to describe with an excellent precision five-particle (or more) states, 
there is still room for different theoretical models. Examples of different models are the chiral 
soliton model of Diakonov, Petrov and Polyakov with a rotational excitation in the pseudoscalar field
\cite{Diakonov1}, 
the anti-triplet diquark model of Jaffe and Wilczek, Karliner and Lipkin
with a p-wave excitation
\cite{Jaffe2,Lipkin}, 
and the heptaquark model of PB and Marques, applied in this paper, 
of Llanes-Estrada, Oset and Mateu, and of Kishimoto and Sato
\cite{Bicudo00,Llanes-Estrada,Kishimoto},
or tri-linear-molecular model, with a quark-antiquark excitation. 
More experimental data on the pentaquarks are necessary to test the different models.

Moreover the observation of the 
$D^{*-}p$(3100) at H1
\cite{H1}
and the observation of double-charmed baryons at 
SELEX 
\cite{SELEX},
and the expected search of double-charmed baryons at COMPASS
\cite{COMPASS}
suggest that many new pentaquarks with one or two heavy quarks may still be 
discovered. Multiquarks are indeed favoured by the presence of several different 
flavours
\cite{Richard,Lutz}. 
In this paper I perform a systematic exploration of pentaquarks with
any possible combination of flavours. For different searches, see
references 
\cite{Zhu:2004za,Zhu:2004xa,Roberts:2004rh,Lu:2005jn,He:2005ey}.

\par
Here, multiquarks are studied
microscopically in a standard quark-model (QM) Hamiltonian. 
Any multiquark state can be formally decomposed in combinations of 
simpler colour-singlet clusters, the baryons and mesons. 
The energy of the multiquark state is computed with the multiquark matrix 
element of the QM Hamiltonian. These matrix elements also
produce the short range interaction of the mesonic or baryonic
subclusters of the multiquark. However, in the case of any pure exotic 
groundstate pentaquark, the short-range interaction can be shown to be repulsive
\cite{Bicudo_KN1,Bicudo_KN2,Bender,Barnes,Hyslop}.
This repulsion agrees with the quenched lattice simulations, which in the range of 
the $\Theta^+$ experimental mass, only identify a parity - system, compatible with 
an open Kaon-nucleon channel. 

To search for attraction, one should also study other cluster-cluster
interactions. For instance in the $N+N$ system, the attractive 
long-range One-Pion-Exchange-Potential and the medium-range Sigma-Exchange 
Potential are crucial for the binding of the deuteron. In a microscopic quark 
Hamiltonian perspective these interactions are equivalent to the coupling 
to a channel with an extra pion ($^3S_1$ quark-antiquark cluster) plus a 
p-wave excitation, and to a second channel with two extra s-wave pions 
respectively. 
Because the long range interaction is vanishing or small, and the medium range 
interaction may be insufficient to provide all the desired attraction, the 
five-quark systems are too unstable to produce narrow pentaquarks.

\par
Nevertheless, one may consider that an s-wave flavour-singlet light 
quark-antiquark pair $l \bar l$ is added to the pentaquark $M$. When the 
resulting heptaquark $M'$ remains bound, it is a state with parity opposite 
to the original $M$ 
\cite{Nowak}, 
where the reversed parity occurs due to the intrinsic 
parity of fermions and anti-fermions. 
The ground-state of $M'$ is also naturally rearranged in an s-wave baryon 
and in two s-wave mesons, where the two outer hadrons are 
repelled, while the central hadron provides stability.
The mass of the heptaquark $M'$ is expected to be slightly lower 
than the exact sum of these standard hadron masses due to the binding energy.
Because the s-wave pion is the lightest hadron, the minimum energy needed to 
create a quark-antiquark pair can be as small as 100-200 MeV. This energy shift 
is lower than the typical energy of 300-600 MeV of spin-isospin or angular 
excitations in hadrons. Therefore, the first excitation of multiquarks is 
quark-antiquark creation. Moreover, the heptaquarks $M'$ may only decay into 
a low-energy p-wave channel (after the extra quark-antiquark pair is 
annihilated), resulting in a very narrow decay width, consistent with the 
observed exotic flavour pentaquarks. The $M'$ is then a linear 
molecular system, light and with a narrow width.

\par
In very recent works this principle was used to 
indicate that the $\Theta^+$(1540) is probably a 
$K \bullet  \pi \bullet  N$ molecule with binding energy of 30 MeV
\cite{Bicudo00,Llanes-Estrada,Kishimoto}, 
and the $\Xi^{--}$(1862) is a $\bar K \bullet  N \bullet  \bar K$ 
molecule with a binding energy of 60 MeV
\cite{Bicudo00,Bicudo02}.
I also suggest that the new positive parity  
scalar  $D_s(2320)$ and axial $D_{s+}(2460)$ are
$\bar K \bullet  D$ and $\bar K \bullet D^*$ multiquarks 
\cite{Bicudo01},
in agreement with the independent models of
Barnes, Close and Lipkin and of Terasaki
\cite{Barnes:2003dj,Terasaki:2003qa,Terasaki:2005ne},
and that the $D^{*-} p$(3100) is consistent with a  
$D^* \bullet  \pi \bullet  N$ 
linear molecule with an energy of 15 MeV above threshold
\cite{Bicudo00,Bicudo03}.
Assuming this description of the presently observed pentaquarks,
I now predict all possible exotic strange, charmed and bottomed 
pentaquarks compatible with the linear-molecule model. 

\par
In Section \ref{sec:cri} the exotic baryon-meson 
short-range s-wave interaction is studied. I find repulsion in 
exotic multiquarks, and attraction in the channels with 
quark-antiquark annihilation. 
In Section \ref{sec:exo} the possible hadrons with exotic pentaquark flavours
are qualitatively studied with an additional quark-antiquark pair.
This study includes pentaquarks 
with any number of strange, charmed and bottomed quarks.
I find that several new exotics may still be observed.
In section \ref{sec:con} the conclusion is presented.
%
%
%
%
\begin{figure}[t]
\caption{
Examples of overlaps of the Resonant Group Method are depicted, 
in (a) the norm overlap for the meson-baryon interaction, 
in (b) a kinetic overlap for the meson-meson interaction,
in (c) an interaction overlap for the meson-meson interaction, 
in (d) the annihilation overlap for the meson-baryon interaction. 
These overlaps are simple matrix elements of operators with a product of four
wave-functions. While the norm overlap (a) contributes  to the denominators of 
eqs. (\ref{intra kernel}) and (\ref{overlap kernel}), the other overlaps 
(b), (c) and (d) contribute to the effective potential $V_{AB}$.
\label{RGM overlaps}}
\epsfig{file=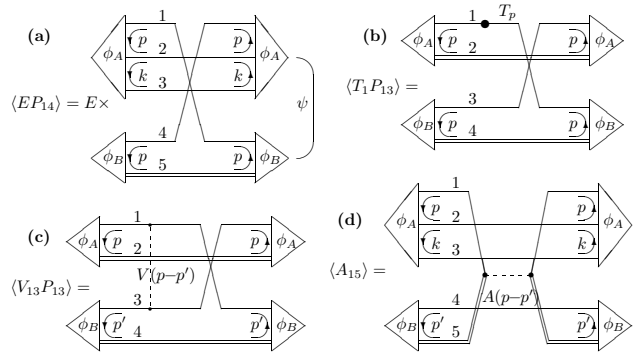,width=8.5cm} 
\end{figure}
%
%

\section{\label{sec:cri}The attraction/repulsion criterion}

\par
The standard QM Hamiltonian is,
\begin{equation}
H= \sum_i T_i + \sum_{i<j} V_{ij} +\sum_{i \bar j} A_{i \bar j} \ , 
\label{Hamiltonian}
\end{equation}
see Fig. ~ \ref{RGM overlaps}. Each quark or antiquark has a kinetic energy $T_i$.
The colour-dependent two-body interaction $V_{ij}$ includes 
the standard QM confining and hyperfine terms,
\begin{equation}
V_{ij}= \frac{-3}{16} \vec \lambda_i  \cdot   \vec \lambda_j
\left[V_{conf}(r) + V_{hyp} (r) { \vec S_i } \cdot { \vec S_j }
\right] \ .
\label{potential}
\end{equation}
The potential of eq. (\ref{potential}) 
reproduces the meson and baryon spectrum with quark and antiquark
bound states (from heavy quarkonium to the light pion mass).
Moreover, the Resonating Group Method (RGM)
\cite{Wheeler}
was applied by Ribeiro, by Oka and Yasaki, and by Toki
\cite{Ribeiro,Oka,Toki} 
to show that in exotic
$N + N$ scattering the quark two-body-potential, together with
the Pauli repulsion of quarks, explains the $N + N$ hard core
repulsion. 
Recently, addressing a tetraquark system with $\pi+\pi$ quantum 
numbers, it was shown that the QM with the quark-antiquark 
annihilation $A_{i \bar j}$ also fully complies with chiral 
symmetry, including the Adler zero and the Weinberg theorem
\cite{Bicudo0,Bicudo1,Llanes-Estrada:2003ha,Bicudo2}. 

\par
The RGM 
\cite{Wheeler}
computes the effective multiquark energy using overlaps, 
or matrix elements with a larger number of wavefunctions, 
of the microscopic quark-quark interactions. The different
examples of overlaps are depicted in Fig. \ref{RGM overlaps}. 
Any multiquark state can be formally decomposed into combinations of 
simpler colour singlets clusters, the baryons and mesons. 
Once the internal energies $E_A$ and $E_B$ of the two hadronic
clusters in the multiquark are accounted for, i. e. , 
\begin{equation}
{\langle \phi_B \phi_A | H \sum_p (-1)^p P |\phi_A \phi_B \rangle
\over
\langle \phi_B \phi_A | \sum_p (-1)^p P |\phi_A \phi_B \rangle
} = E_A+E_B +V_{A \, B} \ ,
\label{intra kernel}
\end{equation}
where $\sum_p (-1)^p P$ is the antisymmetrizer, the remaining energy of 
the multiquark is the effective hadron-hadron potential $V_{A \, B}$.
The meson-baryon or meson-meson potential is computed with the overlap of 
the inter-cluster microscopic potentials
\begin{eqnarray}
V_{\text{bar } A \atop \text{mes } B}
&=& \langle \phi_B \, \phi_A |
-( V_{14}+V_{15}+2V_{24}+2V_{25} )3 P_{14}
\nonumber \\
\nonumber \\
&& +3A_{15} | \phi_A \phi_B \rangle /
\langle \phi_B \, \phi_A | 1- 3 P_{14} | \phi_A \phi_B \rangle
\nonumber \\
V_{\text{mes } A \atop \text{mes } B}
&=& \langle \phi_B \,
\phi_A | (1+P_{AB})[ -( V_{13}+V_{23}+V_{14}+V_{24})  
\nonumber \\
&& \times P_{13} +A_{23}+A_{14} ]| \phi_A \phi_B \rangle 
\nonumber \\
&& / \langle \phi_B \, \phi_A | 
(1+P_{AB})(1-  P_{13}) | \phi_A \phi_B \rangle
\ ,
\label{overlap kernel} 
\end{eqnarray}
where $P_{ij}$ stands for the exchange of particle $i$ with
particle $j$, see Fig. \ref{RGM overlaps}. 
Actually, $V_{A \, B}$ only corresponds to the short range
interaction of the hadrons $A$ and $B$. 
%
%
\begin{table}[t]
\caption{\label{no heavy} 
Masses and decay processes of exotic flavour pentaquarks with no heavy quark.
The method to arrive at this table is described in
Section \ref{sec:exo}. The possible resonances are
divided in classes characterized by the flavour of
light $l=u,d$ or strange $s$ quarks(antiquarks), 
and the number or heavy $H=c,b$ quarks(antiquarks).
}
\begin{ruledtabular}
\begin{tabular}{c|cc}
molecule   
& $~$ mass [GeV] & decay channels \\ \hline
 \multicolumn{3}{l}
{ $ I=1/2, \ ssss \bar l (+3 \, l \bar l) $ : five-hadron molecule}
\\
\hline
\multicolumn{3}{l}
{ $ I=1, \ sssl \bar l  (+2 \, l \bar l) $ : four-hadron molecule}
\\
\hline
\multicolumn{3}{l}
{ $ I=3/2, \ ssll \bar l  (+l \bar l) = s \bar l \bullet   lll \bullet   s \bar l $ }
\\
$ \bar K \bullet   N \bullet   \bar K {\bf = \Xi^{--} } $&  {\em 1.86 } &$ 
\bar K + \Sigma, \, \pi + \Xi
$\\
\hline
\multicolumn{3}{l}
{ $I=2, \ slll \bar l  (+l \bar l) =  s\bar l \bullet   lll \bullet   l \bar l $: pion unbound}
\\
\hline
\multicolumn{3}{l}
{ $I=5/2, \ llll \bar l (+l \bar l) =  l\bar l \bullet   lll \bullet   l \bar l$: pion unbound}
\\
\hline
\multicolumn{3}{l}
{ $I=0, \ llll \bar s (+l \bar l) = l \bar s \bullet   l\bar l \bullet   lll $} 
\\   
$K \bullet   \pi \bullet   N { \bf =\Theta^+ }  $& {\em 1.54 }  &$ 
K + N 
$\\ 
\end{tabular}
\end{ruledtabular}
\end{table}

\par
For the purpose of this paper the details of the potentials in eq. 
(\ref{Hamiltonian}) are unimportant, only their matrix elements matter,
because the effective hadron-hadron interaction can be decomposed
in RGM overlaps. The overlaps are similar to the matrix elements
occuring in the variational method fo the Schr\"odinger equation, where the 
product of two wavefunctions is replaced by a product of four wavefunctions. 
For example the RGM overlap (c) of Fig. \ref{RGM overlaps} is 
identical to the matrix element of the hyperfine potential
times an algebraic number and a factor with the dimension of the 
square of the wave-function. 
The hadron spectrum constrains the matrix element of the
hyperfine potential
\begin{equation}
\langle V_{hyp} \rangle \simeq \frac{4}{3} 
\left( M_\Delta-M_N \right)
\simeq M_{K^*}- M_K  \ .
\label{hyperfine}
\end{equation}
When a light quark is replaced by a heavy quark
\cite{Villate:1992np},
it must also be replaced by 
$\langle{V_{hyp}}_D \rangle \simeq M_{D^*}- M_D$.
The quark-antiquark annihilation potential $A_{i \bar j}$ is 
also constrained when the quark model produces spontaneous 
chiral-symmetry breaking
\cite{Bicudo3}.
The annihilation potential $A$ is present in the $\pi$
Salpeter equation,
\begin{equation}
\left[
\begin{array}{cc}
2 T + V & A \\
A & 2T +V
\end{array}
\right]
\left(
\begin{array}{c}
\phi^+ \\
\phi^-
\end{array}
\right) =
M_\pi
\left(
\begin{array}{c}
\phi^+ \\
-\phi^-
\end{array}
\right) \ ,
\label{pion BS}
\end{equation}
where the $\pi$ is the only hadron with a large negative-energy
wavefunction, $\phi^- \simeq \phi^+$.  In eq. (\ref{pion BS}) the
annihilation potential $A$ cancels most of the kinetic energy and
confining potential $2T+V$. This is the reason why the pion has a very
small mass. From the hadron spectrum and using eq. (\ref{pion BS}),
the matrix elements of the annihilation potential are determined as,
\begin{eqnarray}
\langle 2T+V \rangle_{S=0} &\simeq& {2 \over 3} (2M_N-M_\Delta)
\nonumber \\
\Rightarrow \langle A \rangle_{S=0} &\simeq&- {2 \over 3} 
(2M_N-M_\Delta)
\ ,
\label{sum rules}
\end{eqnarray}
which is correct for the annihilation of $u$ or $d$ quarks,
and nearly correct for the $s$ quark.
%
\begin{table}[t]
\caption{\label{one heavy} 
Masses and decay processes of exotic flavour pentaquarks with one heavy quark. 
The method to arrive at this table is described in
Section \ref{sec:exo}. The possible resonances are
divided in classes characterized by the flavour of
light $l=u,d$ or strange $s$ quarks (antiquarks), 
and the number or heavy $H=c,b$ quarks (antiquarks).
}
\begin{ruledtabular}
\begin{tabular}{c|cc}
linear molecule   
& $~$ mass [GeV] & decay channels \\ \hline
\multicolumn{3}{l}
{ $ I=1/2, Hsss \bar l (+2 \, l \bar l) $ : four-hadron molecule}
\\ 
\hline
\multicolumn{3}{l}
{ $ I=1, Hssl \bar l (+l \bar l)=      s \bar l \bullet  l l H \bullet  s \bar l   $ }
\\
$ \bar K \bullet  \Lambda_c \bullet  \bar K $&$  3.23 \pm 0.03 $&$  \bar K + \Xi_c  , \, \pi +\Omega_c 
$\\
$ \bar K \bullet  \Lambda_b \bullet  \bar K $&$  6.57 \pm 0.03 $&$  \bar K + \Xi_b  , \, \pi +\Omega_b
$\\
\hline
\multicolumn{3}{l}
{ $ I=3/2, Hsll \bar l (+l \bar l)=      s \bar l \bullet  lll \bullet  H \bar l   $}
\\
$  \bar K \bullet  N \bullet  D $&$ 3.25 \pm 0.03 $&$ \bar K + \Sigma_c  , \, D + \Sigma , \, \pi + \Xi_c 
$\\ 
$ \bar K \bullet  N \bullet  D^* $&$  3.39 \pm 0.03 $&$  \bar K + \Sigma_c  , \, D^* + \Sigma , \, \pi + \Xi_c 
$\\ 
$ \bar K \bullet  N \bullet  \bar B $&$ 6.66 \pm 0.03 $&$ \bar K + \Sigma_b  , \, \bar B + \Sigma , \, \pi + \Xi_b

$\\ 
$ \bar K \bullet  N \bullet  \bar B^* $&$  6.71 \pm 0.03 $&$  \bar K + \Sigma_b  , \, \bar B^* + \Sigma , \, \pi + \Xi_b

$\\
\hline
\multicolumn{3}{l}
{ $ I=2, Hlll \bar l (+l \bar l)= l \bar l \bullet  l l l \bullet  H \bar l   $ 
: pion unbound}
\\
\hline  
\multicolumn{3}{l}
{ $ I=1/2, Hlll \bar s  (+l \bar l)=  l \bar s \bullet  l \bar l \bullet  l l H  $ }
\\  
$ K \bullet  \pi \bullet  \Sigma_c $&$ 3.08 \pm 0.03 $&$  K + \Lambda_c , \, K + \Sigma_c  , \, D_s +N
$\\
$ K \bullet  \pi \bullet  \Sigma_b $&$ 6.41 \pm 0.1  $&$  K + \Lambda_b , \, K + \Sigma_b  , \, D_s +N
$\\
\multicolumn{3}{l}
{ $ I=1/2, Hlll \bar s  (+l \bar l)= l \bar s \bullet  H \bar l \bullet  lll  $ }
\\  
$ K \bullet  D \bullet  N $&$  3.25 \pm 0.03 $&$  K + \Lambda_c , \ K + \Sigma_c , \ D_s + N 
$\\
$ K \bullet  D^* \bullet  N $&$ 3.39  \pm 0.03 $&$  K + \Lambda_c  , \ K + \Sigma_c  , \ D_s^* + N 
$\\
$ K \bullet \bar B \bullet  N $&$  6.66 \pm 0.03 $&$  K + \Lambda_b  , \ K + \Sigma_b , \ \bar B_s + N  
$\\
$
K \bullet \bar B^* \bullet  N $&$ 6.71 \pm 0.03 $&$   K + \Lambda_b  , \ K + \Sigma_b , \ \bar B_s^* + N 
$\\
\end{tabular}
\end{ruledtabular}
\end{table}

\par
The annihilation potential only shows up in non-exotic channels, and it 
is clear from eq. (\ref{sum rules}) that the annihilation potential 
provides an attractive (negative) interaction. 
The quark-quark(antiquark) potential is 
dominated by the interplay of the hyperfine interaction of eq. (\ref{hyperfine})
and the quark exchange of eq. (\ref{overlap kernel}).
In s-wave systems with low spin this results in a repulsive interaction. 
The short-range interactions are independent of the details of the 
chiral-invariant quark model that one chooses to consider.
Therefore, I arrive at the attraction/repulsion criterion for groundstate hadrons:
\\
- {\em whenever the two interacting hadrons have quarks (or antiquarks)
with a common flavour, the repulsion is increased by the Pauli principle;
\\
- when the two interacting hadrons have a quark and an
antiquark with the same flavour, the attraction 
is enhanced by the quark-antiquark annihilation}.

\par
For instance, $uud-s \bar u$ is attractive,
and $uud-u \bar s$ is repulsive. In the cases where attraction
is added to repulsion, it turns out that attraction prevails 
when there are more quark-antiquark matchings than quark-quark ones.
For example, $uus-s \bar u$ is attractive whereas $uss-s \bar u$ is repulsive. 
This qualitative rule is confirmed by quantitative computations 
of the short-range interactions of the $\pi , \, N , \, K , \, D , \,
D^* , \, B , \, B^* $ 
\cite{Bicudo00,Bicudo01,Bicudo03,Bicudo02,Bicudo0,Bicudo1,Bicudo2}. 
This rule can also be applied to other baryons. I find for example 
that the $\pi \bullet \Lambda$ short-range interaction is vanishing, 
while the $I=3/2, \, \bar K \bullet \Sigma$ interaction is repulsive.

\begin{table}[t]
\caption{\label{one anti-heavy} 
Masses and decay processes of exotic flavour pentaquarks with one heavy anti-quark.
The method to arrive at this table is described in
Section \ref{sec:exo}. The possible resonances are
divided in classes characterized by the flavour of
light $l=u,d$ or strange $s$ quarks (antiquarks), 
and the number or heavy $H=c,b$ quarks (antiquarks).
}
\begin{ruledtabular}
\begin{tabular}{c|cc}
linear molecule   
& $~$ mass [GeV] & decay channels \\ \hline
\multicolumn{3}{l}
{ $ I=0, ssss \bar H (+3 l \bar l)$ : five-hadron molecule}
\\
\hline
\multicolumn{3}{l}
{ $ I=1/2, sssl \bar H (+2 \, l \bar l)  $ : four-hadron molecule}
\\
\hline
\multicolumn{3}{l}
{ $ I=0, ssll \bar H (+l \bar l) =  l \bar H \bullet  l \bar l \bullet  lss $ }
\\
$  \bar D \bullet  \pi \bullet  \Xi $&$  3.31 \pm 0.03 $&$  \bar D + \Xi  
, \, \bar D_s + \Lambda 
$\\
$  \bar D^* \bullet  \pi \bullet  \Xi $&$  3.45 \pm 0.03 $&$  \bar D^* + \Xi
, \, \bar D^*_s + \Lambda , \, \bar D_s + \Lambda 
$\\
$  B \bullet  \pi \bullet  \Xi $&$  6.73 \pm 0.03 $&$  B + \Xi  
, \, B_s + \Lambda
$\\
$  B^* \bullet  \pi \bullet  \Xi $&$  6.77 \pm 0.03 $&$  B^* + \Xi  
, \, B^*_s + \Lambda , \, B_s + \Lambda
$\\
\hline
\multicolumn{3}{l}
{ $ I=1/2, slll \bar H (+l \bar l) =  l \bar H \bullet  l \bar l \bullet  l l s  $ }
\\
$ \bar D \bullet  \pi \bullet  \Sigma $&$  3.19 \pm 0.03 $&$  \bar D + \Lambda , \, \bar D + \Sigma  , \, \bar D_s +N
$\\
$ \bar D^* \bullet  \pi \bullet  \Sigma $&$  3.33 \pm 0.03 $&$  \bar D^* + \Lambda , \, \bar D^* + \Sigma  , \, \bar D^*_s +N
$\\
$ B \bullet  \pi \bullet  \Sigma $&$  6.60 \pm 0.03 $&$  B + \Lambda , \, B + \Sigma  , \, B_s +N

$\\
$ B^* \bullet  \pi \bullet  \Sigma $&$  6.64 \pm 0.03 $&$  B^* + \Lambda , \, B^* + \Sigma  , \, B^*_s +N

$\\
\multicolumn{3}{l}
{ $ I=1/2, slll \bar H (+l \bar l) =   l \bar H \bullet  s\bar l \bullet  l l l   $ }
\\
$  \bar D \bullet  \bar K \bullet  N $&$  3.25 \pm 0.03 $&$  \bar D + \Lambda , \, \bar D + \Sigma  , \, \bar D_s +N
$\\
$ \bar D^* \bullet  \bar K \bullet  N $&$ 3.39 \pm 0.03 $&$   \bar D^* + \Lambda , \, \bar D^* + \Sigma  , \, \bar D^*_s +N

$\\
$ B \bullet  \bar K \bullet  N $&$ 6.66 \pm 0.03 $&$  B + \Lambda , \, B + \Sigma  , \, B_s +N

$\\
$ B^* \bullet  \bar K \bullet  N $&$  6.71 \pm 0.03 $&$  B^* + \Lambda , \, B^* + \Sigma  , \, B^*_s +N

$\\
\hline
\multicolumn{3}{l}
{ $ I=0, llll \bar H (+l \bar l)=  l \bar H \bullet  l \bar l \bullet  lll $}
\\
$ \bar D \bullet  \pi \bullet  N  $&$  2.93 \pm 0.03 $&$ \bar D+ N  
$\\ 
$ \bar D^* \bullet  \pi \bullet  N {\bf  = \bar D^{*-}p } $& {\em 3.10 } &$ \bar D^*+ N , \, \bar D +N
$\\ 
$ B \bullet  \pi \bullet  N $&$ 6.35 \pm 0.03 $&$ B+N  
$\\ 
$ B^* \bullet  \pi \bullet  N $&$  6.39 \pm 0.03 $&$ B^*+N, B+N 
$\\ 
\end{tabular}
\end{ruledtabular}
\end{table}

\section{\label{sec:exo}Exotic-flavour pentaquarks}

\par 
The exotic pentaquarks containing five quarks only are not 
expected to bind, due to the attraction/repulsion criterion. 
To increase binding we include a light $l \bar l$ 
quark-antiquark pair in the system. 
%
\begin{table}[t]
\caption{\label{two heavy} 
Masses and decay processes of exotic flavour pentaquarks with two heavy quarks.
The method to arrive at this table is described in
Section \ref{sec:exo}. The possible resonances are
divided in classes characterized by the flavour of
light $l=u,d$ or strange $s$ quarks (antiquarks), 
and the number or heavy $H=c,b$ quarks (antiquarks).
}
\begin{ruledtabular}
\begin{tabular}{l|cc}
linear molecule   
& $~$ mass [GeV] & decay channels \\ \hline
 \multicolumn{3}{l}
{ $I=1/2, \ HHss \bar l (+ l \bar l)=H \bar l \bullet lss \bullet H \bar l $}
\\ 
$
D \bullet \Xi \bullet D $&$  5.00 \pm 0.03 $&$ 
D + \Omega_c  , \, \bar K + \Omega_{cc}
$\\ 
$
D \bullet \Xi \bullet D^* $&$  5.14 \pm 0.03 $&$ 
D(D^*) + \Omega_c , \, \bar K + \Omega_{cc} 
$\\ 
$
D^* \bullet \Xi \bullet D^* $&$  5.29 \pm 0.03 $&$ 
D^* + \Omega_c , \, \bar K + \Omega_{cc}
$\\ 
$
D \bullet \Xi \bullet \bar B (\bar B^*) $&$  8.41 \, (8.46) \pm 0.03 $&$  
D + \Omega_b , \, \bar B (\bar B^*) + \Omega_c  
$\\ 
$
D^* \bullet \Xi \bullet \bar B (\bar B^*) $&$  8.56 \, (8.60) \pm 0.03 $&$  
D^* + \Omega_b , \, \bar B (\bar B^*) + \Omega_c 
$\\ 
$
\bar B \bullet \Xi \bullet \bar B (\bar B^*) $&$  11.83(11.87)
 \pm 0.03 $&$ \bar B (\bar B^*) + \Omega_b  , \, \bar K + \Omega_{bb} 
$\\ 
$
\bar B^* \bullet \Xi \bullet \bar B (\bar B^*) $&$ 11.87 (11.92)
 \pm 0.03 $&$\bar B (\bar B^*) + \Omega_b  , \, \bar K + \Omega_{bb} 
$\\ 
 \multicolumn{3}{l}
{ $I=1/2, \ HHss \bar l (+ l \bar l)= s \bar l \bullet l H H \bullet s \bar l   $}
\\ 
$
\bar K \bullet \Xi_{cc} \bullet \bar K $&$  4.52 \pm 0.1  $&$ \bar K + \Omega_{cc}  , \, D + \Omega_c
$\\ 
$
\bar K \bullet \Xi_{cb} ( \Xi_{bb} ) \bullet \bar K $&$  7.77 \, ( 11.02) \pm 0.1  $&$
 \bar K + \Omega_{cb}  ( \Omega_{bb} ) , \, D + \Omega_b 
$\\ 
\hline
 \multicolumn{3}{l}
{ $I=1, \ HHsl \bar l (+ l \bar l)=H \bar l \bullet l l s \bullet H \bar l   $}
\\ 
$
D \bullet \Lambda \bullet D $&$  4.80 \pm 0.03 $&$ D + \Xi_c  , \, \bar K + \Xi_{cc} , \, \pi + \Omega_{cc}
$\\ 
$
D \bullet \Lambda \bullet D^* $&$  4.94 \pm 0.03 $&$ D(D^*) + \Xi_c , \, \bar K + \Xi_{cc} 
$\\ 
$
D^* \bullet \Lambda \bullet D^* $&$  5.08 \pm 0.03 $&$ 
D^* + \Xi_c  , \, \bar K + \Xi_{cc} 
$\\ 
$
D \bullet \Lambda \bullet \bar B (\bar B^*) $&$  8.21 (8.26) \pm 0.03 $&$ 
D + \Xi_b , \, \bar B (\bar B^*) + \Xi_c  
$\\ 
$
D^* \bullet \Lambda \bullet \bar B (\bar B^*) $&$  8.35 (8.40)\pm 0.03 $&$ 
D^* + \Xi_b , \, \bar B (\bar B^*) + \Xi_c  
$\\ 
$
\bar B \bullet \Lambda \bullet \bar B (\bar B^*) $&$  11.62 (11.67) \pm 0.03 $&$ 
\bar B (\bar B^*) + \Xi_b  , \, \bar K + \Xi_{bb}  
$\\ 
$
\bar B^* \bullet \Lambda \bullet \bar B (\bar B^*) $&$  11.67 (11.72) \pm 0.03 $&$ 
\bar B (\bar B^*) + \Xi_b , \, \bar K + \Xi_{bb}  
$\\ 
 \multicolumn{3}{l}
{ $I=1, \ HHsl \bar l (+ l \bar l)=H \bar l \bullet l l H \bullet s \bar l $}
\\
$
\bar B (\bar B^*) \bullet \Lambda_c \bullet \bar K $&$  8.01 (8.06) \pm 0.03 $&$ 
 \bar B (\bar B^*) + \Xi_c  , \, \Xi_{cb} + \bar K
$\\ 
$
D (D^*) \bullet \Lambda_b \bullet \bar K $&$  7.94 (8.08) \pm 0.1  $&$ 
 D (D^*) + \Xi_b   , \, \Xi_{cb} + \bar K
$\\ 
\hline
 \multicolumn{3}{l}
{ $I=3/2, \ HHll \bar l (+ l \bar l)= H \bar l \bullet l l l \bullet H \bar l $}
\\
$
D \bullet N \bullet D $&$  4.62 \pm 0.03 $&$ 
 D + \Sigma_c  , \, \pi + \Xi_{cc}
$\\ 
$
D \bullet N \bullet D^* $&$  4.76 \pm 0.03 $&$ 
D + \Sigma_c , \, D^* + \Sigma_c  
$\\ 
$
D^* \bullet N \bullet D^* $&$  4.91 \pm 0.03 $&$ 
D^* + \Sigma_c  , \, \pi + \Xi_{cc}

$\\ 
$
D \bullet N \bullet \bar B (\bar B^*) $&$  8.04 \, (8.08) \pm 0.03 $&$ 
D + \Sigma_b , \, \bar B (\bar B^*) + \Sigma_c  
$\\ 
$
D^* \bullet N \bullet \bar B (\bar B^*) $&$  8.18 \, (8.22)\pm 0.03 $&$ 
D^* + \Sigma_b , \, \bar B (\bar B^*)  + \Sigma_c  
$\\ 
$
\bar B \bullet N \bullet \bar B (\bar B^*)  $&$  11.45 \, (11.49) \pm 0.03 $&$ 
\bar B ( \bar B^* ) + \Sigma_b , \, \pi + \Xi_{bb}

$\\ 
$
\bar B^* \bullet N \bullet \bar B (\bar B^*)  $&$  11.49 \, (11.54) \pm 0.03 $&$ 
\bar B ( \bar B^* ) + \Sigma_b , \, \pi + \Xi_{bb}
$\\ \hline
 \multicolumn{3}{l}
{ $I=0, \ HHll \bar s (+ l \bar l)= l \bar s \bullet l \bar l \bullet l H H  $}
\\
$ K \bullet \pi \bullet \Xi_{cc} $&$  4.20 \pm 0.1  $&$ K + \Xi_{cc}, \, 
D_s + \Lambda_c
$\\
$ K \bullet \pi \bullet \Xi_{cb} (\Xi_{bb}) $&$  7.45 \, (10.70) \pm 0.1  
$&$ K + \Xi_{cb} (\Xi_{bb}) , \, D_s + \Lambda_c
$\\
 \multicolumn{3}{l}
{ $I=0, \ HHll \bar s (+ l \bar l)= l \bar s \bullet H \bar l \bullet l l H  $}
\\
$ K \bullet D (D^*) \bullet \Lambda_b $&$  8.01 (8.06) \pm 0.1  $&$ K + \Xi_{cb} , 
\, D_s (D_s^*) + \Lambda_b
$\\
$ K \bullet \bar B (\bar B^*) \bullet \Lambda_c $&$  7.94 (8.08) \pm 0.03  $&$ K + \Xi_{cb} , 
\, \bar B_s (\bar B_s^*) + \Lambda_c
$\\
\end{tabular}
\end{ruledtabular}
\end{table}
I now detail the strategy to find the possible linear heptaquark molecules. 
\\
{\bf a)} I consider any heptaquark with the flavour of an exotic pentaquark
and with up to two heavy flavours or anti-flavours. 
Hadrons with three heavy flavour are presently quite hard to produce in the
laboratory. The top quark is excluded because it is too unstable. 
To minimise the short-range repulsion and to increase the attraction of 
the three-hadron system, I only consider pentaquarks with a minimally
exotic isospin, and with low spin. 
%
\begin{table}[t]
\caption{\label{one heavy one anti-heavy} 
Masses and decay processes of exotic flavour pentaquarks with one heavy quark and one heavy anti-quark.
The method to arrive at this table is described in
Section \ref{sec:exo}. The possible resonances are
divided in classes characterized by the flavour of
light $l=u,d$ or strange $s$ quarks (antiquarks), 
and the number or heavy $H=c,b$ quarks (antiquarks).
}
\begin{ruledtabular}
\begin{tabular}{c|cc}
linear molecule   
& $~$ mass [GeV] & decay channels \\ \hline
 \multicolumn{3}{l}
{ $I=0, Hsss \bar H' (+2 \, l \bar l) $ : four-hadron molecule}
\\ 
\hline
 \multicolumn{3}{l}
{ $ I=1/2, Hssl \bar H' (+l \bar l)= lss \bullet H \bar l \bullet l \bar H'$}
\\ 
$ \Xi \bullet \bar D \bullet \bar B (\bar B^*) $&$  8.41 \, (8.46) \pm 0.03 $&$ 
\Xi + \bar B_c (\bar B_c^*) , \,  \Omega_c + \bar B (\bar B^*)   
$\\
$ \Xi \bullet \bar D^* \bullet \bar B (\bar B^*) $&$  8.55 \, (8.60) \pm 0.03 $&$ 
\Xi + \bar B_c (\bar B_c^*) , \,  \Omega_c + \bar B (\bar B^*)  
$\\
$ \Xi \bullet B (\bar B^*) \bullet D $&$  8.41 \, (8.46) \pm 0.03 $&$ 
\Xi + B_c (\bar B_c^*) , \,  \Omega_b + D  
$\\
$ \Xi \bullet B (\bar B^*) \bullet D^* $&$  8.55 (8.60) \pm 0.03 $&$
 \Xi + B_c(\bar B_c^*)  , \,  \Omega_b + D^*  
$\\
\hline
\multicolumn{3}{l}
{ $ I=0, Hsll \bar H' (+l \bar l)= l l H \bullet s \bar l  \bullet l  \bar H' $}
\\ 
$ \Lambda_b  \bullet \bar K \bullet \bar D (\bar D^*)  $&$  7.94 (8.08) \pm 0.1  $&$ 
\Xi_b + \bar D (D^*), \ \Lambda_b + \bar D_s (\bar D_s^*)
$\\ 
$ \Lambda_c \bullet \bar K \bullet B  (B^*) $&$  8.01 (8.06) \pm 0.03 $&$ 
\Xi_c + B (B^*) , \ \Lambda_c + B_s (B_s^*) 
$\\ 
\multicolumn{3}{l}
{ $ I=0, Hsll \bar H' (+l \bar l)= l l s \bullet H \bar l \bullet l \bar H' $}
\\
$ \Lambda \bullet D \bullet B  (B^*)  $&$  8.21 (8.26) \pm 0.03 $&$ \Xi_c + B(B^*)  , \Lambda + B_d(B_d^*)  
$\\ 
$ \Lambda \bullet D^* \bullet B (B^*)    $&$  8.35 (8.40) \pm 0.03 $&$ \Xi_c + B (B^*)  , \Lambda + B_d (B_d^*)  
$\\ 
$ \Lambda \bullet \bar B (\bar B^*) \bullet \bar D  $&$  8.21 (8.26) \pm 0.03 $&$ \Xi_b + \bar D    , \Lambda + \bar D_b (\bar D_b^*)    
$\\ 
$ \Lambda \bullet \bar B ( \bar B^*) \bullet \bar D^*  $&$  8.35 (8.40) \pm 0.03 $&$ \Xi_b + \bar D^*, \Lambda + \bar D_b (\bar D_b^*)
$\\ 
\hline
\multicolumn{3}{l}
{ $ I=1/2, Hlll \bar H' (+l \bar l)= l l H \bullet l \bar l \bullet l \bar H' $}
\\ 
$ \Sigma_c \bullet \pi \bullet B ( B^*)  $&$  7.86 \, (7.91)\pm 0.03 $&$ \Lambda_c (\Sigma_c) + B ( B^*) , \, N + B_d
$\\ 
$ \Sigma_b \bullet \pi \bullet \bar D(\bar D^*)   $&$  7.78 (7.92) \pm 0.1 $&$ \Lambda_b (\Sigma_b) + \bar D (\bar D^*) , \, N + \bar B_c 
$\\ 
\multicolumn{3}{l}
{ $ I=1/2, Hlll \bar H' (+l \bar l)= l l l \bullet H \bar l \bullet l \bar H' $}
\\ 
$ N \bullet D \bullet B (B^*)  $&$  8.04 \, (8.08) \pm 0.03 $&$ \Lambda_c + B (B^*) , N + B_d (B_d^*) 
$\\ 
$ N \bullet D^* \bullet B (\bar B^*)  $&$  8.18 \, (8.22) \pm 0.03 $&$ \Lambda_c + B (B^*) , N + B_d^* (B_d) 
$\\ 
$ N \bullet \bar B (\bar B^*) \bullet \bar D  $&$  8.04 \, (8.08) \pm 0.03 $&$ \Lambda_b + \bar D, N + \bar B_c (\bar B_c^*) 
$\\ 
$ N \bullet \bar B (\bar B^*) \bullet \bar D^*  $&$  8.18 \, (8.22) \pm 0.03 $&$ \Lambda_b + \bar D^*, N + (\bar B_c) \bar B_c^*
$\\ 
\end{tabular}
\end{ruledtabular}
\end{table}
%
\\
{\bf b)} Here the flavour is decomposed in an s-wave system of 
a spin $1/2$ baryon and two 
pseudoscalar mesons, except for the vectors $D^*$ and $B^*$ 
which are also considered. 
\\
{\bf c)} I only consider, as candidates for narrow pentaquarks,
the systems where at least one hadron is attracted by both other ones.
The attraction/repulsion criterion is used to discriminate which hadrons
are bound and which are repelled. 
\\
{\bf d)} In the case of some exotic flavour pentaquarks, only a 
four-hadron-molecule or a five-hadron-molecule would bind. These 
cases are not detailed in the tables, because they are 
difficult to create in the laboratory.
\\
{\bf e)} Moreover, in the particular case where one of the three hadrons 
is a $\pi$, binding is only assumed if the $\pi$ is the central hadron, 
attracted both by the other two ones. The $\pi$ is too light 
to be bound by just one hadron
\cite{Bicudo00}.
\\
{\bf f)} The masses of the bound states with a pion are computed assuming a total 
binding energy of the order of 10 MeV, averaging the binding energy of the
$\Theta^+$ and of the $D^{*-}p$ system in the molecular perspective. The masses of the 
other bound states are computed assuming a total binding energy of the order of 50 MeV, 
averaging the binding energies of the $\Xi^{--}$ and of the new positive-parity $D_S$
mesons. 
\\
{\bf g)}
For a more precise binding energy one would need to include the attractive 
medium-range interaction, the full three-body Fadeev effects, and the coupling to p-wave 
decay channels. Nevertheless all these effects should increase the
binding energy, without changing the heptaquark picture of this paper.
This results in an error bar of $\pm$ 30 MeV for the mass of the multiquark.
When one of the hadrons in the molecule is not listed by the Particle Data Group
\cite{RPP}, 
the hadron mass is extracted from a recent lattice computation
\cite{Mathur},
and the error bar for the mass of the multiquark is $\pm$ 100 MeV.
\\
{\bf h)}
Although three-body decay channels are possible through quark rearrangement,
their observation requires high experimental statistics. 
Only some of the different possible two-body decay processes are detailed here.

\par
The possible narrow exotic-flavour pentaquarks are summarised in Tables  
\ref{no heavy}, 
\ref{one heavy}, 
\ref{one anti-heavy}, 
\ref{two heavy} and
\ref{one heavy one anti-heavy}.
For each flavour the heptaquark structure is produced, and the 
possible molecular states are listed together with the possible two-body decay 
processes. Three-body decays may also be possible, but they are not detailed here.
For isospin $I \neq 0$ the decay processes are shown in a condensed form. 
For instance, the $\Xi^{--}$ observed at NA49
\cite{Alt}
belongs to an iso-quadruplet, and in Table \ref{no heavy} the decay process is
summarised into $\bar K + \Sigma$, where $K$ corresponds to either 
$K^+$ or to $K^0$ while $\bar K$ corresponds either to $K^-$ or $\bar K^0$.
When the isospin is specified, the different members of the quadruplet decay 
respectively to $ K^- + \Xi^-$, to $K^- + \Xi^0$ and $\bar K^0 + \Xi^-$, to 
$K^- + \Xi^+$ and $\bar K^0 + \Xi^0$, and to $\bar K^0 + \Xi^+$.
Moreover, Tables \ref{two heavy} and \ref{one heavy one anti-heavy} have
a very large number of states, and therefore these tables are further condensed
when at least one of the heavy quarks is a bottom quark. 
For instance, $B(B^*)$ means that both $B$ and $B^*$ states should be considered.

\section{\label{sec:con}Conclusion and outlook}

\par
This work has performed a systematic search of exotic-flavour pentaquarks, 
using the linear three-body hadronic-molecule perspective. 
This perspective is the result of standard QM computations
of pentaquarks masses and of hadron-hadron short-range interactions. 
This view, where the quark-antiquark annihilation is crucial, 
explains the difficulty of quenched lattice QCD computations in identifying
the positive parity pentaquark $0^+$.
I only consider pentaquarks with an extra light $l \bar l $ 
quark-antiquark pair (total parity +), minimal exotic isospin, and spin 1/2. 
These are the most favourable conditions to have exotic pentaquarks, 
but nevertheless more exotic states may still be observed.
Because quark models are not fully calibrated yet, the results are
affected by an error bar. 

\par
A large number of new exotic flavour-pentaquarks are predicted in Tables
\ref{no heavy}, 
\ref{one heavy}, 
\ref{one anti-heavy}, 
\ref{two heavy} and
\ref{one heavy one anti-heavy},
together with their decay channels. 
It is interesting to remark that degenerate states occur in Tables \ref{one heavy} 
and \ref{two heavy}, and in Tables \ref{one anti-heavy} and \ref{one heavy one anti-heavy}.

Moreover, some new multiquarks are easier to bind than the presently observed 
exotic pentaquarks. In particular two very promising new exotic-flavour pentaquark 
candidates are the $I=1/2, \ slll\bar (+ l \bar l) c=N \bullet \bar K \bullet D$ 
and $slll\bar (+ l \bar l) c=N \bullet \bar K \bullet D^*$ states. Notice that
the $D_s$(2317) might be a $I=0, \ \bar K \bullet D$ state, the $D_s$(2460) 
might be a $I=0, \ \bar K \bullet D^*$ state with 60 MeV binding energies,
\cite{Bicudo01,Barnes:2003dj,Terasaki:2003qa,Terasaki:2005ne},
and that the $\Lambda$(1405) may also be a $I=0, \ N \bullet \bar K$
with binding energy of 30 MeV (mixed with a $\Sigma \bullet \pi $ state)
\cite{Oset1,Oset2}. This particular pentaquark candidate is thus
expected to have a large binding energy, of the order of 100 MeV,
possibly larger than the binding energy  of the other resonances of its class.  
Moreover the $\pi$, the $\bar K$, the $N$ and the $D^*$
can be detected by most experimental collaborations.
Thus these two resonances, with masses of the order of respectively 3.20 to 3.25 Gev
and 3.34 to 3.39 GeV, decaying into the two-body decays listed in
Table \ref{one anti-heavy} and also in three-body decays, say in
$\bar D (\bar D^*) + \pi + \Sigma$ are excellent candidates for new
pentaquarks. If one of these new resonances is observed, it will
provide an excellent evidence for hadronic molecules. 

\par
The quantitative computation of the masses, decay rates and sizes
of some of the proposed heptaquarks will be done elsewhere. 
The most relevant contributions that remain to be included in this framework
are the attractive medium-range interaction,
the full three-body Fadeev effects, and the coupling to p-wave decay channels.

\acknowledgments
I am grateful to Katerina Lipka, Achim Geiser, Paula Bordalo and Pedro Abreu for 
discussions on the possibility to detect new exotic pentaquarks. This work is 
devoted to encourage the experimental search for new exotic multiquarks.


\end{document}